# Topological Hall Effect and Skyrmion-like Bubbles at a Charge-transfer Interface

*Zhi Shiuh Lim, Changjian Li, Zhen Huang, Xiao Chi, Jun Zhou, Shengwei Zeng, Ganesh Ji Omar, Yuan Ping Feng, Andrivo Rusydi, Stephen John Pennycook, Thirumalai Venkatesan, Ariando Ariando**

Dr. Z. S. Lim, Dr. C. Li, Dr. Z. Huang, Dr. S.W. Zeng, G. J. Omar, Prof. T. Venkatesan, Assoc. Prof. A. Ariando
NUSNNI-NanoCore, National University of Singapore, Singapore 117411
email: ariando@nus.edu.sg

Dr. Z. S. Lim, Dr. Z. Huang, Dr. X. Chi, Dr. J. Zhou, Dr. S.W. Zeng, G. J. Omar, Prof. Y. P. Feng, Prof. T. Venkatesan, Assoc. Prof. A. Rusydi, Assoc. Prof. A. Ariando
Department of Physics, National University of Singapore, Singapore 117542

Dr. X. Chi, Assoc. Prof. A. Rusydi
Singapore Synchrotron Light Source (SSLS), National University of Singapore, 5 Research Link, Singapire 117603

Dr. C. Li, Prof. S. J. Pennycook, Prof. T. Venkatesan
Department of Materials Science and Engineering, National University of Singapore, Singapore 117575

**Abstract:**

Exploring exotic interface magnetism due to charge transfer and strong spin-orbit coupling has profound application in future development of spintronic memory. Here, the emergence, tuning and interpretation of hump-shape Hall Effect from a $CaMnO_3$/$CaIrO_3$/$CaMnO_3$ trilayer structure are studied in detail. The hump signal can be recognized as Topological Hall Effect suggesting the presence of Skyrmion-like magnetic bubbles; but the debated alternative interpretation where the signal being an artefact between two cancelling Anomalous Hall Effect loops is also discussed. Firstly, by tilting the magnetic field direction, the evolution of Hall signal suggests transformation of the bubbles' topology into a more trivial kind. Secondly, by varying the thickness of $CaMnO_3$, the optimal thicknesses for the hump signal emergence are found, suggesting a tuning of charge transfer fraction. Using high-resolution transmission electron microscopy, a stacking fault is also identified, which distinguishes the top and bottom $CaMnO_3$/$CaIrO_3$ interfaces in terms of charge transfer fraction and possible interfacial Dzyaloshinskii-Moriya Interaction. Finally, a spin-transfer torque experiment revealed a low threshold current density of ~$10^9$ A/m$^2$ for initiating the bubbles' motion. This discovery opens a possible route for integrating Skyrmions with antiferromagnetic spintronics.

The key advantages to be offered by the antiferromagnetic spintronics field are storing information with immunity to external magnetic field perturbation, and ultrafast (~THz) switching dynamics[1]. Recently, this field sparked worldwide research interest when reliable all-electrical manipulation (via spin-orbit torque (SOT)) and detection (via anisotropic magnetoresistance (AMR)) of antiferromagnetic states was achieved in CuMnAs[2]. On the other hand, the abundant ferromagnetic Skyrmions currently known are promising candidates as information-storing bits in the proposed “racetrack memory” design[3]. Skyrmion racetrack memories[4] are expected to operate at ultralow current density[5, 6] and thus energy saving due to Skyrmions’ intrinsic ability to evade defects-pinning[7]. Among the efforts in building such realistic devices, a few crucial challenges must be overcome: Firstly, Skyrmions should be robustly nucleated or deleted at the WRITE-head by a localized current injection[8], or more preferably by a current-less electric field tuning of the magnetic energy landscape[9]. Secondly, when moving driven by current, the Skyrmion Hall Effect (SkHE)[10, 11] is detrimental because the transverse deflection would lead to their annihilation at sidewalls especially at high speed. Several strategies for suppressing the gyromagnetic Magnus force have emerged, namely by stabilizing Skyrmions in ferrimagnetic materials[12] or synthetic antiferromagnetic multilayers coupled by Ruderman–Kittel–Kasuya–Yosida (RKKY) interaction[13], albeit an actual antiferromagnetic Skyrmion remains elusive. Thirdly, Skyrmions detection at the READ-head should be performed by a compact Tunnelling Non-collinear Magnetoresistance (TNcMR)[14] mechanism. Notably, the first and second aspects mentioned above pose some contradictions, i.e.: a metallic racetrack is required for current-driven Skyrmions motion, yet electric field gating for Skyrmions nucleation would require an insulator with low carrier density to be effective. To bridge this gap, researchers would turn attention to multifunctional oxide heterostructures in search of compromising properties.

In perovskite oxides, studies on charge-transfer driven interfacial physics have grown equally fruitful. The “polar-discontinuous” combinations (e.g. $LaAlO_3/SrTiO_3$) yielded confined high-mobility two-dimensional electron gas (2DEG) with magnetism and even superconductivity[15]. Whereas in nonpolar combinations, the driving force of charge transfer has been understood by an initial O2p band alignment before contact and follows the trend of energy gap between O2p and valence band ($E_{O2p}$-$E_F$)[16]; which is distinct from the early Schottky-Mott rule[17] emphasizing work-function difference. Consistent to this concept, charge transfer from $Ir^{4+}$ to $Mn^{4+}$ has been experimentally verified in $SrMnO_3/SrIrO_3$ superlattices[18], and yielded interface ferromagnetism, perpendicular magnetic anisotropy (PMA) and Anomalous Hall Effect (AHE). Besides, owing to the strong spin-orbit coupling (SOC) contributed by $SrIrO_3$, Neel-type Skyrmion is also predicted to exist in $La_{1-x}Sr_xMnO_3/SrIrO_3$ superlattice by a Monte-Carlo simulation[19], which is very valuable since such system is semiconducting and removes the contradiction mentioned above. In this work, we attempt to reduce the complexity of superlattice into a trilayer structure of $CaMnO_3$(3uc)/$CaIrO_3$(19uc)/$CaMnO_3$(3uc), abbreviated as Ca-$Mb_3I_{19}Mt_3$. In addition to AHE, we also observe the emergence of hump-shape Hall features in temperature range of 10-50 K (**Figure 1c**) that resembles Topological Hall Effect (THE), suggesting presence of Skyrmion-like bubbles. By various examinations such as tilting magnetic field direction, tuning $CaMnO_3$ thicknesses (Mb and Mt), atomic survey and current-driven dynamics, our experimental results shall present valuable insights to the community.

**Results:**

The mechanism of THE is well-understood – as conduction electrons hopping across non-coplanar magnetic textures, strong Hund’s coupling ($J_H$) between the electrons’ spin ($\hat{\boldsymbol{s}}$) and the gradually varying local moment ($\hat{\boldsymbol{m}}$) causes the electrons to gain a real-space Berry phase, leading to transverse deflection[20, 21]. This effect requires a net topological charge $Q = \frac{1}{4\pi}\sum_{ijk} 2\tan^{-1}\frac{\hat{\boldsymbol{m}}_k\cdot(\hat{\boldsymbol{m}}_i\times\hat{\boldsymbol{m}}_j)}{1+\hat{\boldsymbol{m}}_i\cdot\hat{\boldsymbol{m}}_j+\hat{\boldsymbol{m}}_j\cdot\hat{\boldsymbol{m}}_k+\hat{\boldsymbol{m}}_k\cdot\hat{\boldsymbol{m}}_i}$ among triads of local moments or equivalently $Q = \frac{1}{4\pi}\int \hat{\boldsymbol{m}}\cdot\left(\frac{\partial\hat{\boldsymbol{m}}}{\partial x}\times\frac{\partial\hat{\boldsymbol{m}}}{\partial y}\right)$ dxdy, and has been observed in manganites[21, 22], B20 compounds[23], pyrochlore frustrated antiferromagnets[24], ultrathin $SrRuO_3$ systems[25, 26], and topological insulators[26-28]. Since it is well-known that a typical magnetic phase evolution involves stripe-domains at zero-field, Skyrmions at intermediate regime, and collinear ferromagnet at large field, $\rho_{THE}$(H) is expected to adopt a hump-shape[26, 29]. While the hysteretic saturated collinear ferromagnetic background of where Skyrmions are embedded will contribute a square $\rho_{AHE}$(H) loop. This is the "AHE+THE" picture.

However, in recognition of the recent debates[30], we caution that the data in Figure 1c can also be interpreted as a partial cancellation between two AHE loops with opposite signs and different coercive fields ($H_c$), which can be simulated by two pairs of Langevin functions: $\mathcal{L}_{12}(\mathrm{H}) = \mathrm{A}_{12}\left\{\coth\left[\frac{g\mu_o\mu_B J_{12}}{k_B T}\left(H \pm H_{c1,2}\right)\right] - \frac{k_B T}{g\mu_o\mu_B J_{12}\left(H \pm H_{c1,2}\right)}\right\}$ (**Figure S1a**). We can infer the non-equivalent top and bottom CMO/CIO interfaces to be the possible sources of opposite-sign AHE loops, such that the hump emergence only hosts trivial collinear domains instead of Skyrmions-like bubbles. To clarify the superexchange mechanism in $Mn^{4+}/Ir^{4+}$ combination, we show a sketch of single-ion density-of-states (DOS) to illustrate the two possible charge transfer pathways into the $e_g^{\uparrow}$ and $t_{2g}^{\downarrow}$ bands of $Mn^{4+}$ that are in proximity in energy (**Figure 1a**). Both the $\mathrm{Ir5d_{J_{eff}=1/2} - O2p_{x,y} - Mn3d_{xz,yz}}$ (π-π) and $\mathrm{Ir5d_{J_{eff}=1/2} - O2p_{x,y} - Mn3d_{3z^2-r^2}}$ (π-σ) transfer pathways are viable (**Figure 1b**), the former is expected to be weak due to low orbital overlap while the latter is enabled by octahedral tilt. This proximity is known from O K edges of X-ray absorption (XAS) in bulk $CaMnO_3$ as a huge peak at ~528.8 eV[31]. Several *ab-initio* calculations indicated that the Mn $e_g^{\uparrow}$ orbital is dominantly filled after charge transfer[32, 33], consistent to our XAS result (**Figure S2b**) suggesting that $e_g^{\uparrow}$ is slightly lower than $t_{2g}^{\downarrow}$. This results in a ferromagnetic Mn-O-Mn double-exchange along xy-plane via the $d_{x^2-y^2}$ orbitals, but an antiparallel Ir-O-Mn superexchange involving the $d_{xz,yz}$ and $d_{3z^2-r^2}$ orbitals along z-direction. Considering the larger density-of-states (DOS($E_F$)) due to heavier effective masses in $t_{2g}$ bands compared to $e_g$, we expect the spin-polarization, $P_s = \frac{DOS_{\uparrow}(E_F)-DOS_{\downarrow}(E_F)}{DOS_{\uparrow}(E_F)+DOS_{\downarrow}(E_F)} < 0$, leading to a negative-sign $\rho_{AHE}$ loop (**Figure 1d**, top). This can be understood from the semiclassical intrinsic transverse velocity[34]: $\boldsymbol{v}_{i\perp,\uparrow\downarrow}(\boldsymbol{k}) = \mp\frac{e}{\hbar}\boldsymbol{\mathcal{E}}_{\parallel} \times \boldsymbol{\Omega}_{i,z}(\boldsymbol{k})$ where $\boldsymbol{\Omega}_i(\boldsymbol{k}) = \boldsymbol{\nabla}_{\boldsymbol{k}} \times \langle u_i(\boldsymbol{k})|\boldsymbol{\nabla}_{\boldsymbol{k}}|u_i(\boldsymbol{k})\rangle$ is the k-space Berry curvature at the $i$-th band and $\boldsymbol{\mathcal{E}}_{\parallel}$ is the linear electric field. Hence, the AHE current density is $\boldsymbol{j}_{\perp} = -e\sum_{i,\uparrow\downarrow}\int_{BZ}\frac{d^3k}{(2\pi)^3} f_D\big(E(k)\big)\, v_{i\perp,\uparrow\downarrow}(k) \approx P_s\frac{ne^2}{\hbar}[\boldsymbol{\mathcal{E}}_{\parallel} \times \langle\boldsymbol{\Omega}_z\rangle]$, where $n$, $f_D$ and $\langle\ldots\rangle$ are total carrier density, Fermi-Dirac distribution and average. Note that such picture of negative-sign $\rho_{AHE}$ loop and antiparallel $M_{Mn}$-$M_{Ir}$ moments alignment agree well with J. Nichols' result[18] and our our first-principle calculation on superlattice (**Figure S4c**). Conversely, if the charge-transfer fraction is reduced, lowering $E_F$ and shut the $t_{2g}^{\downarrow}$ conduction channel, we expect $P_s>0$ (**Figure 1d, bottom**), together with the suppression of xy-plane ferromagnetic Mn-O-Mn into canted antiferromagnetic. Such magnetic phase evolution has been understood via a Hamiltonian $\widetilde{\mathcal{H}} = -t\cos\left(\frac{\beta}{2}\right)\sum_{<ij>}\left(a_{i\uparrow}^{\dagger}a_{j\uparrow} + \mathrm{h.c.}\right) + \sum_{<ij>}J_{ex}\left(\widehat{\boldsymbol{m}}_i \cdot \widehat{\boldsymbol{m}}_j\right)$, which simplifies into an energy equation $E = -4x|t|\cos\left(\frac{\beta}{2}\right) + J_{ex}\cos(\beta)$ to be minimized, where $t$, $\beta$, $J_{ex}$ and $x$ are the hopping integral, angle deviating from ferromagnetic alignment, superexchange coupling and charge-transfer fraction respectively[33]. For $CaMnO_3$ featuring larger octahedral tilts (smaller $t$) and stronger $J_{ex}$ than that of $SrMnO_3$, a higher critical charge-transfer fraction ($x_c = \frac{2J_{ex}}{|t|}$) is required to achieve the ideal ferromagnetic alignment[33]. In turn, the $P_s>0$ scenario when $M_{Mn,canted} > |M_{Ir}|$ would then be more likely to occur, although it has been neglected thus far. In short, we recognize here that the "Bi-AHE" scenario, represented by the negative-sign $\mathcal{L}_2$ and positive-sign $\mathcal{L}_1$, cannot be ruled out.

In the magnetization-temperature (*M-T*) curves (**Figure 2a**), there are interestingly three obvious kinks indicating magnetic phase transitions. We ascribe the lowest $T_1$~65 K to the $T_N$ of antiferromagnetic $Mn^{4+}$–$O^{2-}$–$Mn^{4+}$ superexchange, which is ubiquitous in both in-plane and out-of-plane fields. The intermediate $T_2$~105 K is the $T_C$ of $Mn^{(4-\delta)+}$–$O^{2-}$–$Ir^{(4+\delta)+}$ superexchange with PMA since it only exists at H||[001]; while the highest $T_3$~185 K could be the onset of $Ir^{4+}$ spin nematic/liquid order[35] due to its proximity with $Mn^{4+}$ near the interfaces. Besides, the magnetization-field (*M-H*) loops (**Figure 2b**) are almost temperature independent and very narrow, not showing $H_c$ comparable to that of AHE loops; and correspondingly the magnetoresistance (MR) in **Figure 2f** are negative parabolic but hysteretic "butterfly loops" are absent. Putting aside the humps, the presence of strong AHE signal but small saturation-magnetization ($M_{sat}$) implies a dominant antiferromagnetic $Mn^{4+}$ in CMO, while the narrow $H_c$ could originate from the magnetized $Ir^{4+}$ moments in proximity to $Mn^{4+}$. The z-component of enhanced interfacial $Mn^{(4-\delta)+}$ canted moment due to the charge transfer is almost undetectable in magnetometry but is available to break time-reversal symmetry (TRS) and contribute a net Berry curvature.

To shed light on the anisotropic magneto-transport, we further performed rotation of magnetic field as illustrated in **Figure 2c**. The linear magnetoresistance (AMR) (**Figure 2f-g**) suggests a trend of anisotropy $K_{[011]}<K_{[001]}<K_{[010]}$ in a sequence of increasingly easier magnetic axis (see supporting information). More insights emerge when such rotation configuration is applied on Hall Effect measurement (**Figure 2d-e**). At 30 K and ***H***||[001], the extraordinary part of Hall Effect $\rho_{xy}$-$R_oB$ can be decomposed following either the "AHE+hump" or the "Bi-AHE" schemes. Details of fittings are given in Figure S1a, and we can perceive that the magnitudes of $\mathcal{L}_2$ and hump are proportional. As shown in **Figure 2d**, the net AHE is seen to increase concomitantly with the diminishing of humps within the range of $30^o<\theta<60^o$; conversely, $\mathcal{L}_1$ stays constant although $\mathcal{L}_2$ starts diminishing at $\theta>30^o$. Both the net AHE and $\mathcal{L}_1$ then vanish fast above $\theta>60^o$ because eventually the Hall Effect cannot be measured with an in-plane field. To understand the rising net AHE after $\theta>30^o$, we draw analogy to the topology transformation once observed in several dipolar- (demagnetization) stabilized bubble systems, namely: (Sc,Mg)-doped $Ba_3Fe_{12}O_{19}$[36] and $La_{0.825}Sr_{0.175}MnO_3$[37]. In those systems, stripe domains dominated at zero-field as usual, while Bloch-type Skyrmions ($Q$=1) with mixed clockwise and counterclockwise helicities ($Q_h=\pm\pi/2$)[38] existed at small magnetic fields of ~0.1-0.2 T. Upon tilting the field towards in-plane direction, Bloch-type Skyrmions are transformed into the “type-II” bubbles with $Q$=0 since they are just made of two counter-propagating Bloch lines. This transformation event may suppress THE and enhance AHE, since the type-II bubbles are equivalent to stripe domains. It is also reasonable from Ginzburg-Landau framework[39], where three in-plane helical magnons with mutual $120^o$ wave-vectors in superposition would create a triangular Skyrmion-lattice at small out-of-plane field, but an in-plane field would disturb their balance, transforming Skyrmions back into stripes. Correspondingly in **Figure 2e**, both $H_{c,AHE}$ and $H_{c1}$ increase following the ~1/cos($\theta$) trend, yet the hump peak field ($H_{hp}$) and $H_{c2}$ remain constant with θ up to their abrupt disappearance at $\theta=60^o$. This shows a stark difference between the net AHE ($\mathcal{L}_1$) and hump ($\mathcal{L}_2$) in terms of their anisotropy and origin.

Next, we investigate into thickness dependence of the Hall components, presented in the “AHE+hump” format since its fitting and decomposition are more straightforward. First, assisted by topography surveys (**Figure 3a**), we are confident that the CMO film growth mode is in layer-by-layer with minimal roughness, while the CIO is in Stranski-Krastanov growth mode with tendency to roughen at large thickness, and 19uc is within the acceptable thickness range. In **Figure 3b** (left panel), removing the top CMO (Ca-$Mb_3I_{19}Mt_0$) retains the AHE (with Anomalous Hall angle $\Theta_{AH}$~0.00145), yet suppresses the humps. If following the "AHE+THE" interpretation, we may assume that the top CMO/CIO interface contributes a Dyzaloshinskii-Moriya interaction (DMI) for Neel-type Skyrmions; or if following the "Bi-AHE" interpretation via $\mathcal{L}_{1,2}$, we may assign $\mathcal{L}_2$ for the top interface and $\mathcal{L}_1$ for the bottom. Either way, such distinction between the two interfaces warrants a detailed atomic survey (**Figure 4**). By tuning the top CMO thickness (Mt) at constant bottom CMO thickness (Mb=3uc) (**Figure 3c**), we observed a relatively constant AHE, yet the hump signal reaches a peak at Mt~4uc. On the other hand, tuning Mb at Mt=3uc (**Figure 3d**) produced peaks at Mb~3uc for both the apparent AHE and hump signals. It is possible to explain the absence of hump or $\mathcal{L}_2$ at the four endpoints. While the reason at regime ① is the absence of the top interface, regime ② (large Mt) can be understood as exceeding the Thomas-Fermi screening length ($l_{TF}$) of charge transfer, resulting in a lack of confinement and lower charge transfer density below a certain threshold. We may estimate $l_{TF} = \sqrt{\frac{\varepsilon_0 \varepsilon_r k_B T}{e^2 n_c}}$ ~8uc at 30 K, approximately agreeing with **Figure 3c**, where $n_c = \left(\frac{1}{4\alpha_0}\right)^3 = \left(\frac{e^2 m_e^*}{16\pi\varepsilon_0\varepsilon_r\hbar^2}\right)^3$ from Mott criterion[40], $\alpha_0$ is the effective Bohr radius, $\varepsilon_r$~150 and $m_e^*$~4.3$m_0$ for $CaMnO_3$[41]. On other other hand, at large CMO thickness and low charge transfer fraction, it is also reasonable to expect that the criteria of Skyrmion-like bubbles formation are not fulfilled, i.e. either $\frac{D}{\sqrt{J_{ex}K_u}} < \frac{4}{\pi}$ for DMI-stabilized Skyrmions[42] due to $J_{ex}$ being too strongly antiferromagnetic, or $\frac{2K_u}{\mu_0 {M_{sat}}^2} < 1$ for dipolar-stabilized bubbles[43] due to lack of strong out-of-plane uniaxial anisotropy $K_u$. While regime ③ corresponding to Ca-$Mb_0I_{19}Mt_3$ is trivial: direct deposition of CIO on $LaAlO_3$ (001) results in island growth mode, severely suppresses the charge transfer at all interfaces and thus all extraordinary Hall components vanish. Lastly at regime ④ (large Mb), the argument of exceeding $l_{TF}$ similar to regime ② is also applicable for the bottom CMO, albeit no roughening occurs.

In **Figure S1b-c**, we show that reducing the CIO thickness in the trilayer structure results in a sign-reversal of the Ordinary Hall Effect (OHE) gradient ($R_o$) from negative to positive, accompanying a linear enhancement of net AHE. The $R_o$ sign-reversal is not surprising, since the bulk-like CIO in perovskite polymorph is known as a near-compensated semimetal ($n_e$=$n_h$) with electron and hole pockets residing at different k-points with comparable mobilities[44]. As shown in **Figure 3e-f**, we fabricated $[CMO_{m\text{-}uc}+CIO_{m\text{-}uc}]_{x20/m}$ superlattices to validate the competition between $\mathcal{L}_{1,2}$ controlled by charge transfer fraction. Indeed, a clear sign-reversal of $\rho_{AHE}$ loops from [m=1]-SL (negative) to [m=2]-SL (positive) is observed down to low temperature. A comparison of the charge-transfer fraction is evidenced from the superlattices' XAS Mn $L_{2,3}$ edges peak shifts, O K edges spectral weight transfer, and ferromagnetic $T_c$ (**Figure S2a-d**). This is consistent to our central concept (Figure 1a) that the reduced charge transfer limited by less interface density per unit volume in [m=2]-SL would yield a positive-sign $\rho_{AHE}$ loop. Furthermore in [m=2]-SL and Ca-$Mb_3I_{19}Mt_3$, a thermal excitation at elevated temperature would shift $E_F$ higher and reach the $P_s$<0 conduction channel, resulting in a temperature-driven $\rho_{AHE}$ sign-reversal from positive to negative (~60-70 K). Yet such temperature-driven sign-reversal is never observed in [m=1]-SL, justifying the argument that $e_g^{\uparrow}$ has a lower band edge than $t_{2g}^{\downarrow}$.

In Figure 4, we gain further insight from distinguish the top and bottom interfaces of Ca-$Mb_3I_{19}Mt_3$ in terms of atomic structures and bond angles, by employing cross-sectional scanning transmission electron microscopy (STEM). As shown in **Figure 4a-b**, while the LAO(001) substrate has an AO-termination, the bottom CMO adjacent to LAO(001) is not a perfect $ABO_3$ perovskite – a stacking fault of two adjacent CaO-CaO planes is found, which resembles a portion of $A_3B_2O_7$ member of the Ruddlesden-Popper series. The regional and randomly distributed nature of such stacking fault is evident by comparing **Figure 4a,c**, but is completely absent in the top CMO. This causes the bottom CIO/CMO interface to have a mixed termination and more diffuse than the top CMO/CIO interface. Besides, by analysing the atomic positions, the B-O-B bond angle is revealed to become smaller (bent) with increasing distance away from the substrate (**Figure 4d**). This indicates suppression of the films' octahedral tilt by close proximity to the cubic-lattice LAO(001) substrate[45, 46], and to sign conflict between $t_{2g}$ and $e_g$ orbital wavefunctions[47]; while the top CMO is unaffected, making a bond angle of ~160$^o$ consistent to unstrained CMO[46].

Finally, the Skyrmion dynamics driven by spin-transfer torque (STT) described by the Thiele's equation[7, 48] forms an important aspect in developing Skyrmion racetrack memories. The dissipative and the Magnus force terms respectively contribute the parallel ($v_{sk,||}$) and transverse ($v_{sk,\perp}$) Skyrmion velocity components. Upon de-pinning with current density exceeding a threshold $j_c$, Skyrmions in parallel motion to the drift velocity of spin-polarized electrons $\boldsymbol{v}_e$ would suppress the Topological Hall force[6, 49] by $\boldsymbol{F}_{\mathrm{THE}} = (e/2)(\boldsymbol{v}_e - \boldsymbol{v}_{sk}) \cdot \frac{h}{e\pi r_{sk}^2}\hat{\boldsymbol{z}}$ due to the reducing relative velocity factor. In **Figure 5a**, our Ca-$Mb_3I_{19}Mt_3$ trilayer produced a clear onset with $j_c$~2x10$^9$ A/m$^2$ ($I$~600 µA), which is still high compared to the ultralow benchmark of ~10$^6$ A/m$^2$ set by B20 compounds Bloch-type Skyrmions[5, 6]. Following methods from Ref. [6], the bubbles' velocity components can be estimated by $v_{sk,||}(J) = -\left|\frac{e}{q^e}\right| \frac{j\Delta\rho_{\mathrm{THE}}(j) R_0(j \ll j_c)}{\rho_{\mathrm{THE}}(j \ll j_c)}$ and $v_{sk,\perp} = \frac{j_c}{\sqrt{j^2 - j_c^2}} v_{sk,||}$ (**Figure 5b**). Changing the measurement current mode from continuous to a 20-ms pulse-width produces the similar trend of diminishing $\rho_{\mathrm{THE}}$, yet blurs the $j_c$ onset. Such blurring is probably due to repetitive transition of the bubbles' motion between "creep" and "steady-flow" regimes[10, 50]. Logically, high-current measurement leads to heat dissipation. We expect $H_c$ of AHE be very sensitive to any temperature change; yet Figure 5a shows that $H_c$ remains constant at the diminishing $\rho_{\mathrm{AHE}}$ and $\rho_{\mathrm{THE}}$. We may relate the rate of temperature change $\frac{\partial T}{\partial t} \sim \frac{j^2 \rho_{xx}}{C_P}$ to thermal power via linear resistivity ($\rho_{xx}$) and molar heat capacity ($C_P$). Comparing to typical metallic SOT experiments of ferromagnets where Joule heating is usually severe, our $\rho_{xx}$~$10^{-6}$ Ω.m is ~$10^{2}$ times (higher), $j_c$ is ~$10^{-1}$-$10^{-2}$ times (lower), and $C_P$ (for ceramics) is ~$10^{1}$ times (higher), resulting in $\frac{\partial T}{\partial t}$ to be ~$10^{-1}$-$10^{-3}$ times (lower). This justifies that Joule heating is minimal in the present oxide system.

**Discussions:**

Combining the stacking fault, suppressed bond-bending and mixed termination at the bottom CMO, the bottom CIO/CMO interface can be inferred to suffer from a combination of three possible suppressions as compared to the top CMO/CIO interface, i.e.: charge transfer fraction, Rashba-type DMI and $K_u$. Hence the top interface is more privileged to contribute negative-sign AHE, or support magnetic Skyrmion-like bubbles. However, we note that there is no one "smoking-gun" experimental strategy that can unambiguously distinguish the "AHE+THE" and "Bi-AHE" interpretations without controversy. For such thin film heterostructures where separation from substrates is difficult, a highly spin-sensitive scanning-probe tool for Skyrmions imaging such as the nitrogen-vacancy (NV) magnetometry[51] or scanning nanoSQUID-on-tip[52] would be invaluable in future work. Nevertheless, proving Skyrmion existence does not invalidate the possibility of two sources of k-space Berry curvatures (for AHE) within the same sample, as shown by the Cr-doped $(Bi_{1-x}Sb_x)_2Te_3$ bilayers case[27]. These materials may be broadly classified into "systems with strongly spin-textured bands"[53], which requires more sophisticated theoretical framework and is out-of-scope at present. In conclusion, this work presents the discovery of interesting THE signal evolution at an innovative oxide antiferromagnetic insulator/paramagnetic semimetal interfacial charge-transfer system, suggesting the stabilization of Skyrmion-like magnetic bubbles. This scheme provides an opportunity to access various material combinations across the Periodic Table that may potentially integrate Skyrmion application with antiferromagnetic spintronics, to be unearthed in the near future.

**Methods:**

By using pulsed laser deposition (PLD) at temperature 680$^{o}$C and laser fluence 2.5 J/cm$^2$, $CaMnO_3$ and $CaIrO_3$ films were grown at oxygen ambience of 200 mTorr and 25 mTorr respectively for trilayers, but commonly 75 mTorr for superlattices, followed by cooling at 250 Torr of oxygen to remove oxygen vacancy. Electrical transport and magnetometry were measured in QD Physical Properties Measurement System (PPMS) and Superconducting Quantum Interference Device Vibrating Sample Magnetometer (SQUID-VSM) respectively. Electron Yield XAS was measured at Singapore Synchrotron Light Source (SSLS). Cross sectional High Angle Annular Dark Field (HAADF) and Annular Bright Field (ABF) STEM were performed with a JEOL ARM200F microscope equipped with an ASCOR aberration corrector. Topography scans were done with a Park-Systems NX-10 Atomic Force Microscope (AFM). In Figure 3a, one piece of LAO(001) substrate was cleaved into multiple fragments so as to preserve the same miscut angle and maintain consistency in roughness, while individual layers in a particular sample were grown without breaking vacuum.

**Acknowledgements:**

This research is supported by the Agency for Science, Technology and Research (A*STAR) under its Advanced Manufacturing and Engineering (AME) Individual Research Grant (IRG) (A1983c0034), the National University of Singapore (NUS) Academic Research Fund (R-144-000-403-114), and the Singapore National Research Foundation (NRF) under the Competitive Research Programs (CRP Award No. NRF-CRP15-2015-01).

**Conflict of Interest:**

The authors declare no conflicting interest.

**Figures:**

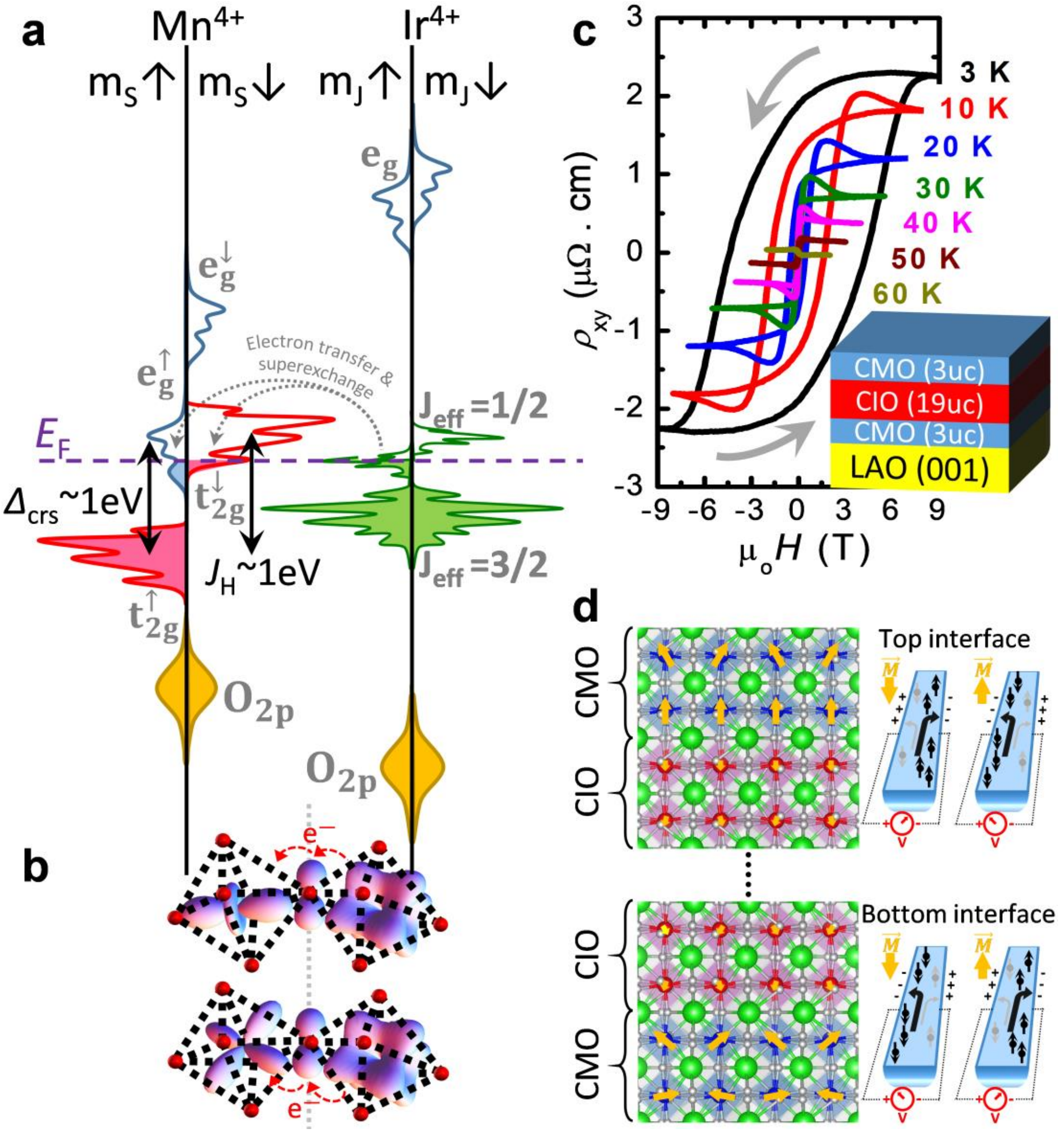

**Figure 1. (a) Fundamental magnetic properties and Hall data in Ca-Mb$_3$I$_{19}$Mt$_3$.** Sketch of single-ion DOS of $Mn^{4+}$ and $Ir^{4+}$ in perovskite B-site after contact and magnetic superexchange. Before contact, O2p bands are aligned but $E_F$ are misaligned (not shown). **(b)** Orbital sketches illustrating the two possible transfer paths in an octahedral-tilted perovskite environment. **(c)** Hall data obtained from the Ca-Mb$_3$I$_{19}$Mt$_3$ structure. **(d)** Left: a sketch illustrating the possible Mn and Ir moments alignment (yellow arrows) at the top and bottom interfaces respectively. Right: the corresponding AHE mechanisms that support the “Bi-AHE” interpretation.

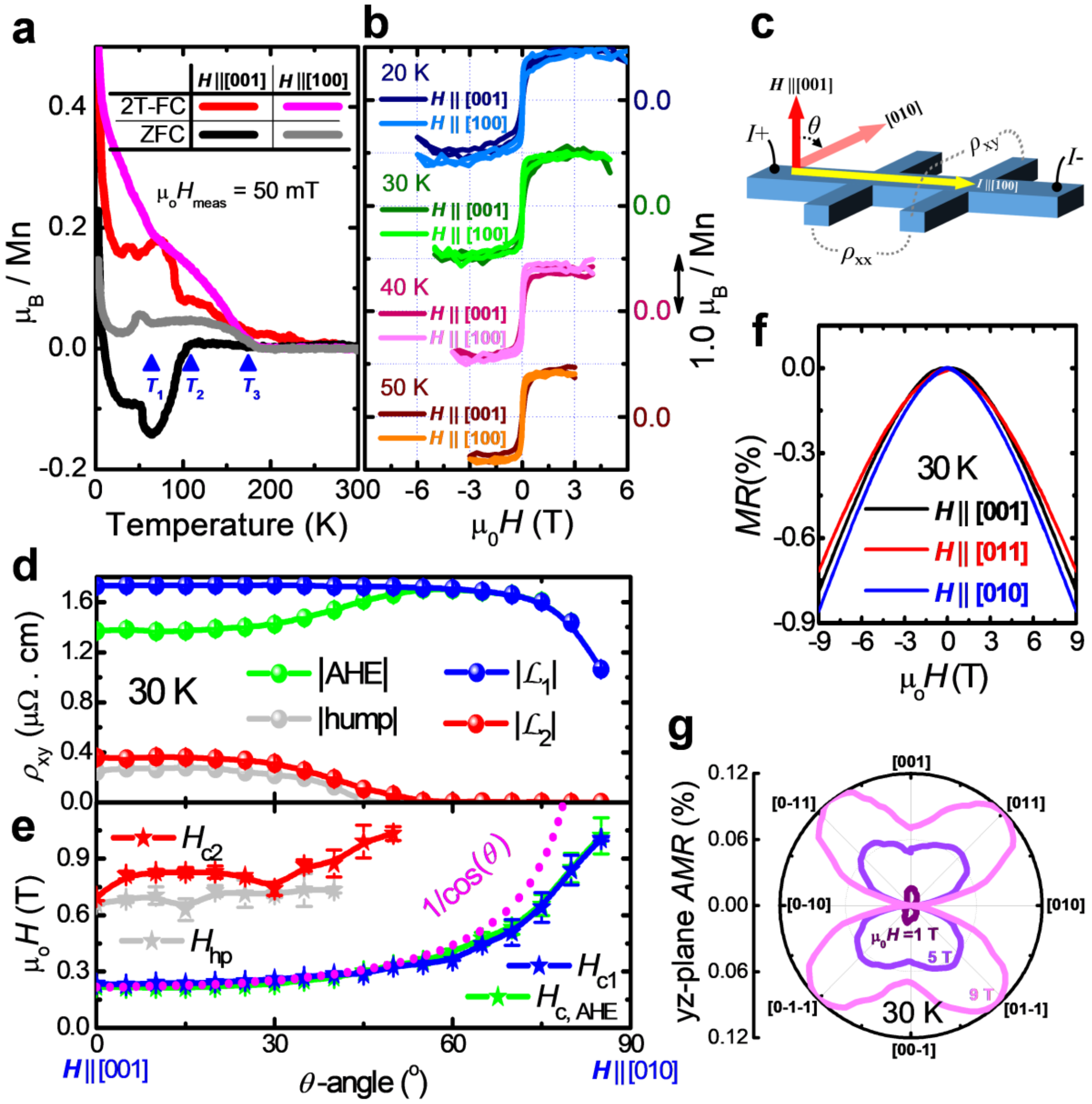

**Figure 2. Magnetometry and anisotropic magneto-transport of Ca-$Mb_3I_{19}Mt_3$. (a)** 2 T-field-cooled (FC) and zero-field-cooled (ZFC) *M-T* and **(b)** *M-H* curves with data shifted vertically for clarity. **(c)** Schematic of $\theta$-dependent Hall measurement. **(d)** $\theta$-dependence of extraordinary Hall Effect decomposed into respective components following the two debated interpretations. **(e)** $\theta$-dependence of $H_{c,AHE}$, $H_{hp}$ and $H_{c1,2}$. The dotted line shows the expected $1/\cos(\theta)$ trend. **(f)** $MR(\mathrm{H}) = \frac{R_{xx}(\mathrm{H})-R_{xx}(H=0)}{R_{xx}(H=0)}$x100% for various magnetic field directions. **(g)** $\mathrm{AMR}(\theta) = \frac{R_{xx}(\theta)-R_{xx}(H||[010])}{R_{xx}(H||[010])}$x100% at various fields.

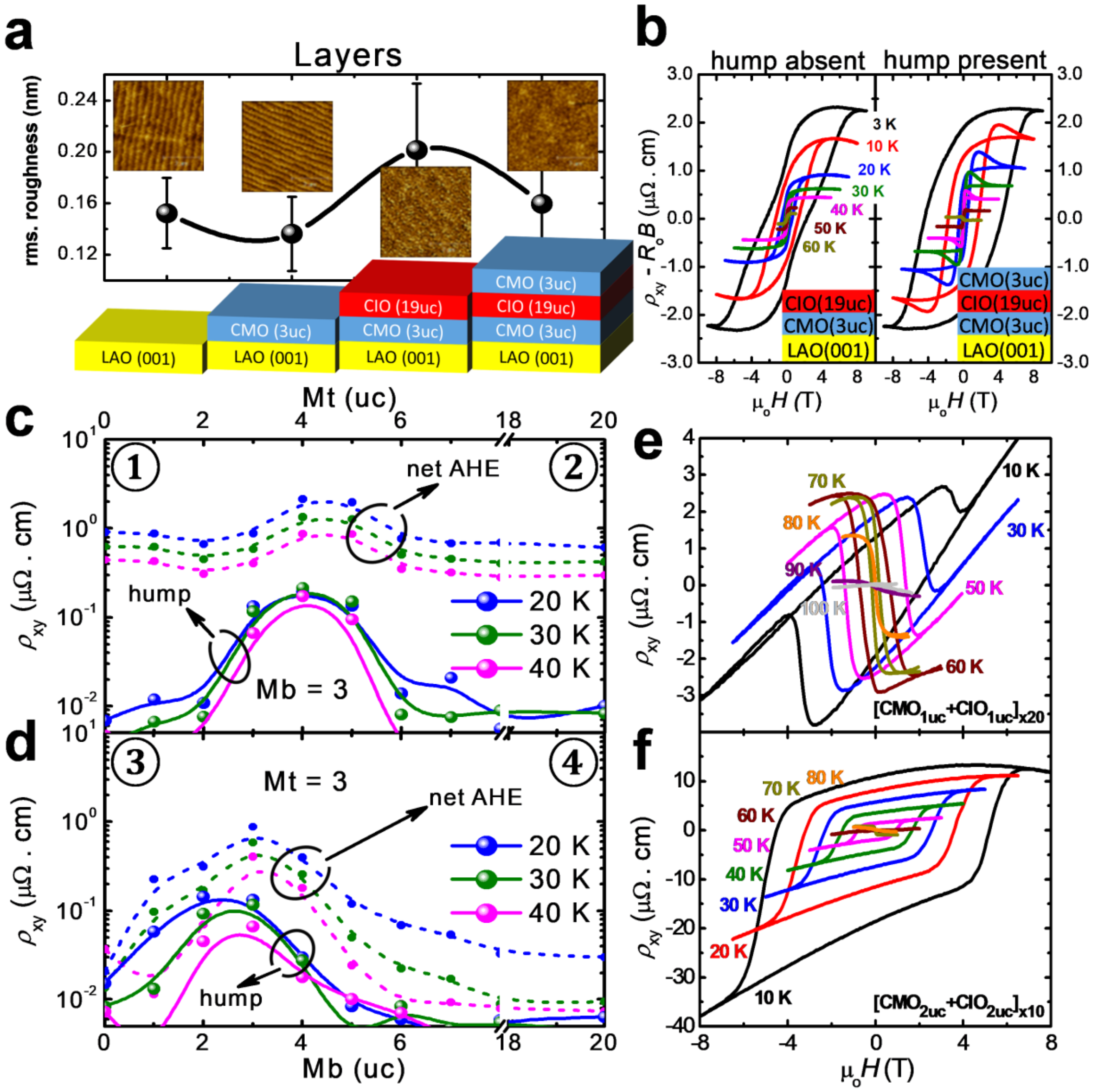

**Figure 3. Tuning the layer thickness of trilayer and superlattices. (a)** Layer-dependent surface roughness with topography attached. **(b)** Comparison between the Hall Effect of Ca-$Mb_3I_{19}Mt_0$ and Ca-$Mb_3I_{19}Mt_3$. Hall Effect variation by tuning the **(c)** top and **(d)** bottom CMO thickness of Ca-$Mb_3I_{19}Mt_3$. Hall Effect of **(e)** [m=1]-SL and **(f)** [m=2]-SL showing opposite-sign AHE loops.

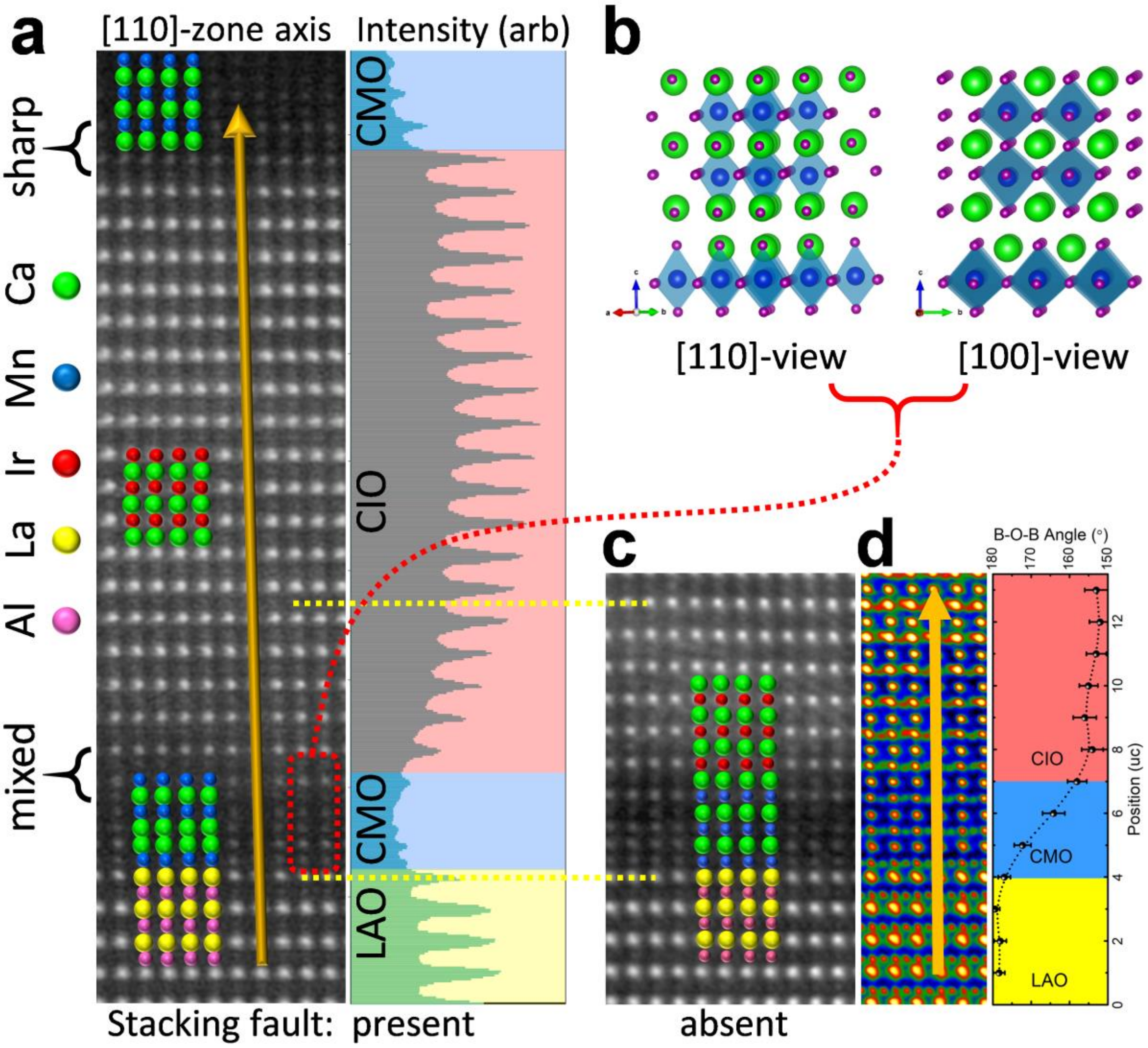

**Figure 4. STEM study on Ca-$Mb_3I_{19}Mt_3$ along [110]-zone axis. (a)** High-angle Annular Dark-field (HAADF) image (left) with element-specific intensity (right). The bracketed region is identified with a stacking fault that can be understood in **(b)**, while 'mixed' refers to termination. **(c)** Image of another region of the same sample similar to (a) but without stacking fault. **(d)** B-O-B bond angle plotted with distance (right panel), extracted from an Annular Bright-field (ABF) image (left panel). Ir-O-Ir bond angles at higher position are not provided due to difficulty in analyzing blurry image. Brown arrows in (a), (d) indicate the selected atoms for intensity and bond angle analyses.

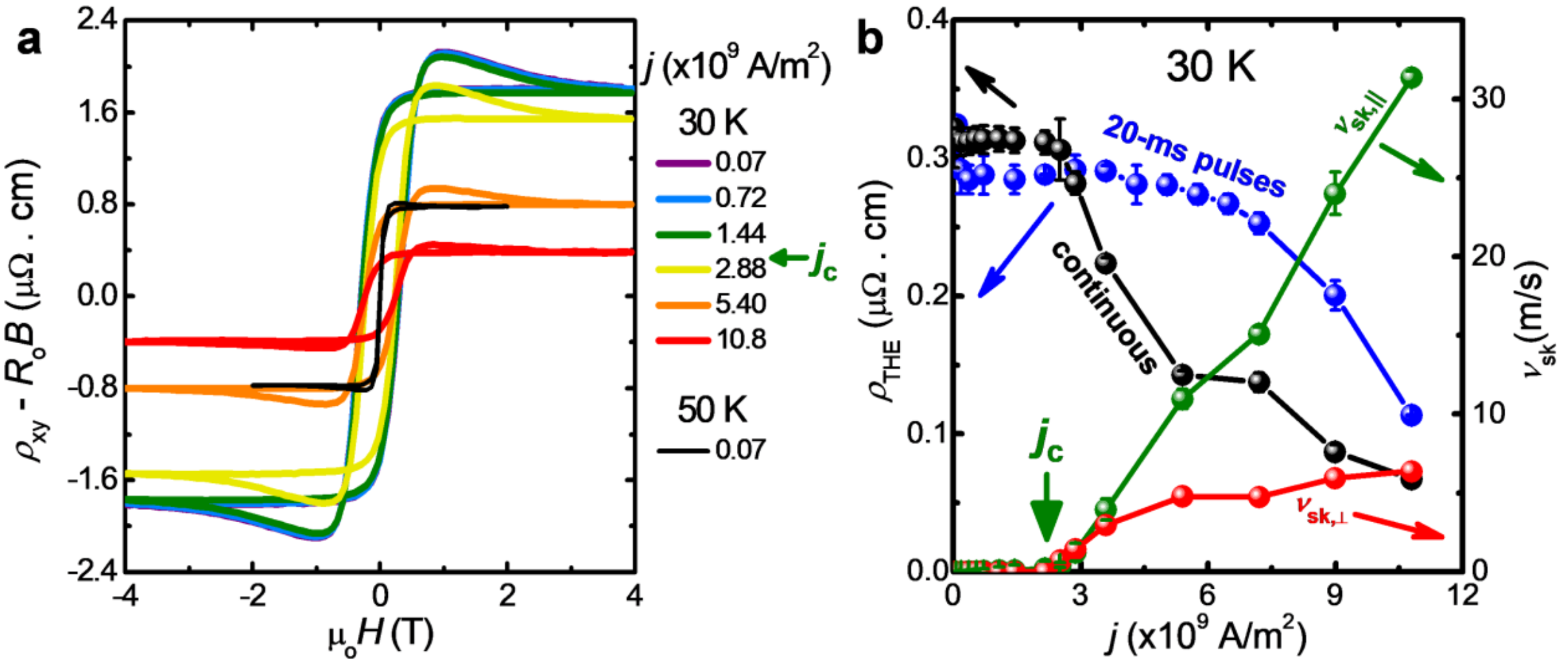

**Figure 5. Current-driven dynamics of magnetic bubbles. (a)** Hall Effect loops with various current densities at 30 K. The 50 K loop at low current with very narrow $H_c$ is shown for comparison to indicate the irrelevance of Joule heating. **(b)** Peak THE (hump signal) variations with increasing current density, under the "continuous" (exacted from (a)) and "20 ms-pulse" current modes. Estimated Skyrmion velocity components from the "continuous" mode curve are plotted together.

# Supporting Information

**Topological Hall Effect and Skyrmion-like Bubbles at a Charge-transfer Interface**

*Zhi Shiuh Lim, Changjian Li, Zhen Huang, Xiao Chi, Jun Zhou, Shengwei Zeng, Ganesh Ji Omar, Yuan Ping Feng, Andrivo Rusydi, Stephen John Pennycook, Thirumalai Venkatesan, Ariando**

Dr. Z. S. Lim, Dr. C. Li, Dr. Z. Huang, Dr. S.W. Zeng, G. J. Omar, Prof. T. Venkatesan, Prof. Ariando
NUSNNI-NanoCore, National University of Singapore, Singapore 117411
email: phyarian@nus.edu.sg

Dr. Z. S. Lim, Dr. Z. Huang, Dr. X. Chi, Dr. J. Zhou, Dr. S.W. Zeng, G. J. Omar, Prof. T. Venkatesan, Prof. A. Rusydi, Prof. Ariando
Department of Physics, National University of Singapore, Singapore 117542

Dr. X. Chi, Prof. A. Rusydi
Singapore Synchrotron Light Source (SSLS), National University of Singapore, 5 Research Link, Singapire 117603

Dr. C. Li, Prof. S. J. Pennycook, Prof. T. Venkatesan,
Department of Materials Science and Engineering, National University of Singapore, Singapore 119077

1. Bi-AHE fittings for the $\theta$-dependent Hall Effect Analyses

**Figure S1a** shows the $\theta$-dependent extraordinary Hall data and their decomposition into humps and (net) AHE following the "AHE+hump" scheme, or into $\mathcal{L}_{1,2}(H)$ following the "Bi-AHE" scheme. The "AHE+hump" fitting is straightforward. The net AHE component is the data up to saturation at large field, and the remaining antisymmetric hump signal, which decays to zero at large field, is obtained by subtracting the net AHE away from the original data. On the other hand, the $\mathcal{L}_{1,2}(H)$ fitting could be arbitrary due to information loss. To assume a reasonable guideline, we assign the AHE loop at $\theta$=60$^{o}$, which appears to be the highest and without humps, as the maximum $\mathcal{L}_1$. This way, $\mathcal{L}_2$ exists up to $\theta$~60$^{o}$ and can be obtained by subtracting $\mathcal{L}_1$ away from the original data. At $\theta$>60$^{o}$, $\mathcal{L}_2$ is zero and not shown.

2. Further discussions on anisotropic magnetoresistance (AMR) measurements

In main text Figure 2f-g, we see a fourfold symmetry with θ-rotation, where the [001] out-of-plane direction could be interpreted as the easy-axis of the ferromagnetic top CMO/CIO interface due to adequate charge transfer; while the [010] in-plane direction could be the easy-axis of the canted antiferromagnetic bottom CIO/CMO interface due to low charge transfer. This way, the [011] is naturally the hard-axis. Note that the measurement configuration (yz-plane θ-rotation) in main text Figure 2c eliminates the SOC-assisted s-d scattering mechanism[1] in AMR, but does not eliminate the Spin Hall Magnetoresistance (SMR)[2]. However, from Ref. [3], the two schemes of out-of-plane to in-plane magnetic field rotation (yz-plane $\theta$-rotation) and (xz-plane $\gamma$-rotation) usually yield the same result, suggesting that both the SOC-assisted s-d scattering and SMR are negligible compared to magnetic anisotropy in our system.

3. CIO thickness variation in CMO/CIO/CMO trilayer

Following the two-carrier model for a compensated semimetal which is a good description for $CaIrO_3$, the ordinary Hall Effect (OHE) collapses into a simpler linear form[4], hence $R_0 =$ $\frac{\partial \rho_{xy}}{\partial B}\Big|_{B\to\infty} = \frac{1}{en}\left(\frac{\mu_h-\mu_e}{\mu_h+\mu_e}\right)$. As shown in **Figure S1b-c**, it is obvious that the magnitudes of net AHE increase with the sign-reversal of $R_o$, although the trend of hump is less obvious. Logically, we expect the part of CIO away from interfaces remain near compensated (bulk-like) but the interfacial CIO within its Thomas-Fermi screening length is significantly hole-doped due to the charge transfer. Hence, at the interface, the holes would dominate over electrons and mainly contributes the AHE signal, and $\mu_e$ could be neglected. Yet, in actual measurement of the trilayer, both the mentioned parts are measured together, forming the total $R_o$ by a simple summation of Hall resistivity and linear conductivity (in parallel circuit) from both parts. Hence, $R_o = \frac{f}{en_{sm}}\left(\frac{\mu_{h,sm}-\mu_{e,sm}}{\mu_{h,sm}+\mu_{e,sm}}\right) + \frac{(1-f)}{en_{hd}}$ , and $\sigma_{xx}(B=0) = f n_{sm} e(\mu_{h,sm} + \mu_{e,sm}) + (1-f) n_{hd}\mu_{hd} e$, where subscript "sm" and "hd" denote "semimetallic" and "hole-doped" respectively, and $f$ is the fraction proportional to the CIO's thickness. We can expect $\mu_{e,sm} > \mu_{h,sm}$ due to the slightly lighter electron's effective mass. This way, the first term of $R_o$ is a negative constant, and $R_o$ sign-reversal occurs following the increasing CIO thickness. Yet, it is reasonable to assume that AHE is only contributed by the interface, hence, the linear scaling relationship seen in Figure S1c is simply reduced to $\rho_{AHE} \propto R_{o,hd} = \frac{C}{en_{hd}}$ (neglecting the semimetallic part), and the relevant $\rho_{xx}(B=0) = \frac{1}{n_{hd}\mu_{hd}e}$ . This way, $\sigma_{AHE} = \frac{\rho_{AHE}}{\rho_{AHE}^2+\rho_{xx}^2(B=0)} \approx \frac{\frac{C}{en_{hd}}}{\left(\frac{C}{en_{hd}}\right)^2+\left(\frac{1}{n_{hd}\mu_{hd}e}\right)^2} = \frac{en_{hd}k\mu_{hd}^2}{\mu_{hd}^2C^2+1}$. Here, it is straightforward that $C$ has the dimension of magnetic field and is ~4 T at 20 K, ~3 T at 30 K, and ~2 T at 40 K. From the supplementary data of Ref. [3], we see that the carrier mobilities are no more than $5\times10^{-3}$ $m^2/V/s$ at these temperatures, hence $\mu_{hd}^2C^2 \ll 1$. This would place our system approximately in the category of bad metal hopping "dirty limit" ($\sigma_{AHE} \sim \mu_{hd}^{1.6} \sim \sigma_{xx}^{1.6}$) with coexistence of both intrinsic and side-jump mechanisms[5]. While the opposite regime $\mu_{hd}^2C^2 \gg 1$ corresponding to the moderately dirty limit ($\sigma_{AHE} \sim \mu_{hd}^0$) is not applicable here. The fast exponential increase of $\rho_{\mathrm{THE}} \propto n^{-8/3}$ as observed in lightly-doped $Ce_xCa_{1-x}MnO_3$ thin films[6], which is the Topological Hall Effect (THE) in the weak coupling regime enhanced by strong correlation and increase in electron's effective mass with reducing $Ce^{4+}$-doping, is also not observed here.

4. Analyses of X-ray Absorption (XAS) spectra

The Mn $L_{2,3}$-edge and O K-edge XAS spectra of pure $CaMnO_3$ film and superlattices at 300 K are shown in **Figure S2a-b**. First, comparing Mn $L_{2,3}$-edges (Figure S2a, inset) which are excitations from the $Mn2p_{1/2}$ and $2p_{3/2}$ core levels to the empty Mn3d valence states, a gradual redshift of peaks can be observed in the sequence from $CaMnO_3$, [m=2]-SL to [m=1]-SL. This indicates that their Mn2p core level electrons have reducing binding energy, implying a gradual reduction of oxidation state of the Mn-species, verifying the charge transfer from $Ir^{4+}$ to $Mn^{4+}$.

Next, we analyse the O K-edge spectra (Figure S2b) for bulk-like $CaMnO_3$, $CaIrO_3$, [m=1] and [m=2] CMO-CIO superlattices, which are excitations from the O1s core level to the empty O2p-Mn3d and O2p-Ir5d hybridized states. We should note: $Mn^{4+}$ is Jahn-Teller inactive and the fully-filled Mn3d $t_{2g}^{\uparrow}$ orbitals do not contribute any O K edge signal. Hence, the peaks are labelled in the table below:

| | | | | |
|---|---|---|---|---|
| Ⓐ | O1s → O2p-Mn3d $e_g^{\uparrow}$ | | | |
| Ⓑ | O1s → O2p-Mn3d $t_{2g}^{\downarrow}$ | | ❶ (A) | O1s → O2p-Ir5d $t_{2g}^{J_{\mathrm{eff}}=1/2}$ |
| Ⓒ | O1s → O2p-Mn3d $e_g^{\downarrow}$ | | ❷ (B) | O1s → O2p-Ir5d $e_g$ |
| Ⓓ,Ⓔ | O1s → O2p-Ca4d | | ❸ (C) | O1s → O2p-Ca4d |
| Ⓕ,Ⓖ,Ⓗ | O1s → O2p-Mn4s/4p | | 🅓 | O1s → O2p-Ir6s/6p |

The near-overlapping peaks Ⓐ and Ⓑ[7, 8] play the key role of stabilizing AHE sign-reversal discussed in this paper. This feature of $CaMnO_3$ is not to be confused with that of $LaMnO_3$ where peaks Ⓐ and Ⓒ are split, since $Mn^{3+}$ with half-filling of the $e_g^{\uparrow}$ band is Jahn-Teller active, and peak Ⓑ is also pushed further up energetically due to increased Hund's coupling ($J_H$)[7, 9]. Peaks Ⓓ, Ⓔ, Ⓕ, Ⓖ, Ⓗ, 🅒, and 🅓 will not be subjected to further analyses. Comparing between CMO, [m=2]-SL and [m=1]-SL, we see reducing peak height at Ⓑ due to increasing occupancy (less empty) at Mn3d $t_{2g}^{\downarrow}$, evidencing charge transfer.

5. <u>Magnetometry for superlattices</u>

Magnetization-temperature (M-T) curves in **Figure S2c-d** show that there is a distinct $T_{C1}$ at ~100 K (for [m=1]-SL and ~80 K for [m=2]-SL, below which $Mn^{(4-\delta)+}$ moments order ferromagnetically. Above $T_{C1}$, the "spin-nematic/liquid" order of $Ir^{(4+\delta)+}$ persists up to >300 K. Perpendicular magnetic anisotropy (PMA) exists in both superlattices. The higher $T_{C1}$ of [m=1]-SL than [m=2]-SL also proves its higher percentage of charge transfer, similar to J. Nichols' result[10]. Since the CIO spacers in the superlattices are thin, the magnetic moments are fully coupled, sharing the same coercive field, hence no hump signal is observable in the Hall Effect of these superlattices (main text Figure 3e-f) although small cancellation may exist.

6. <u>Computational techniques</u>

First-principle calculations were done by Density Functional Theory (DFT) based on the Vienna *Ab-initio* Simulation Package (VASP)[11] with the Local-density Approximation (LDA) for the exchange-correlation functional[12] and the frozen-core all-electron projector-augmented wave (PAW) method[13] for the electron-ion interaction. Mott-Hubbard interaction (U) and Spin-orbit Coupling (SOC) are added to improve accuracy of results, specifically U =

2 eV for Ir-site[14] and U = 5 eV for Mn-site[15]. The cutoff energy for the plane wave expansion was set to 520 eV. For all present cases, a 6×6×3 k-point grid for Brillouin zone sampling is found to be sufficient. All the atoms are allowed to relax until the forces were smaller than 0.01 eV/Å.

The calculated band structures, lattice parameters and Bader charges of bulk perovskite $CaIrO_3$ and $CaMnO_3$ are shown in **Figure S3**, agreeing with earlier publications. The metastable perovskite $CaIrO_3$ polymorph is calculated by LDA-PAW+U+SOC yielding a paramagnetic semimetal ground state[14], this is not to be confused to the more stable post-perovskite $CaIrO_3$ structure which is a spin-orbit Mott insulator[16]. Whereas $CaMnO_3$ is calculated by LDA-PAW+U yielding a G-type antiferromagnetic Mott insulator ground state with Ising-like $\pm$2.712 $\mu_B$/Mn, and an indirect bandgap of 1.46 eV[15, 17].

A supercell of $[CaIrO_3$ (2uc)/$CaMnO_3$ (2uc)$]_{x\infty}$ superlattice is built using 4 Mn, 4 Ir, 8 Ca and 24 O atoms, and the equilibrium atom positions and bond angles after relaxation are shown in **Figure S4a**. In the superlattice, all Mn moments are enhanced compared to the bulk $CaMnO_3$, while Mn and Ir moments are antiparallel and pointing towards the <111> axis. In the orbital-resolved density-of-states (DOS) plotted in **Figure S4b**, $e_g$ orbitals have large DOS at Fermi level ($E_F$). Comparing the Bader charges (**Figure S3f, S4c**) of Mn representing the number of electrons in the 3rd atomic valence shell (~$3s^2 3p^6 3d^3$), there is an increase of 0.11 from the bulk $CaMnO_3$ to the Ca-$M_2I_2$-SL, implying that the charge transfer from Ir to Mn is around $0.11e^-$. This is less than that of (extrapolated) $[SrMnO_3$ (2uc)/$SrIrO_3$ (2uc)$]_{xn}$ superlattice reported in Ref. [[10]] (~$0.35e^-$), probably due to larger octahedral tilt (smaller Mn-O-Ir bond angle) in the present work with smaller $Ca^{2+}$ at the A-site compared to $Sr^{2+}$. Nevertheless, ferromagnetic coupling between Mn moment still exists since it fulfills the critical charge transfer fraction $x_c \sim 0.11$ according to Bhowal's result[18].

**Supporting References:**

**Supporting Figures:**

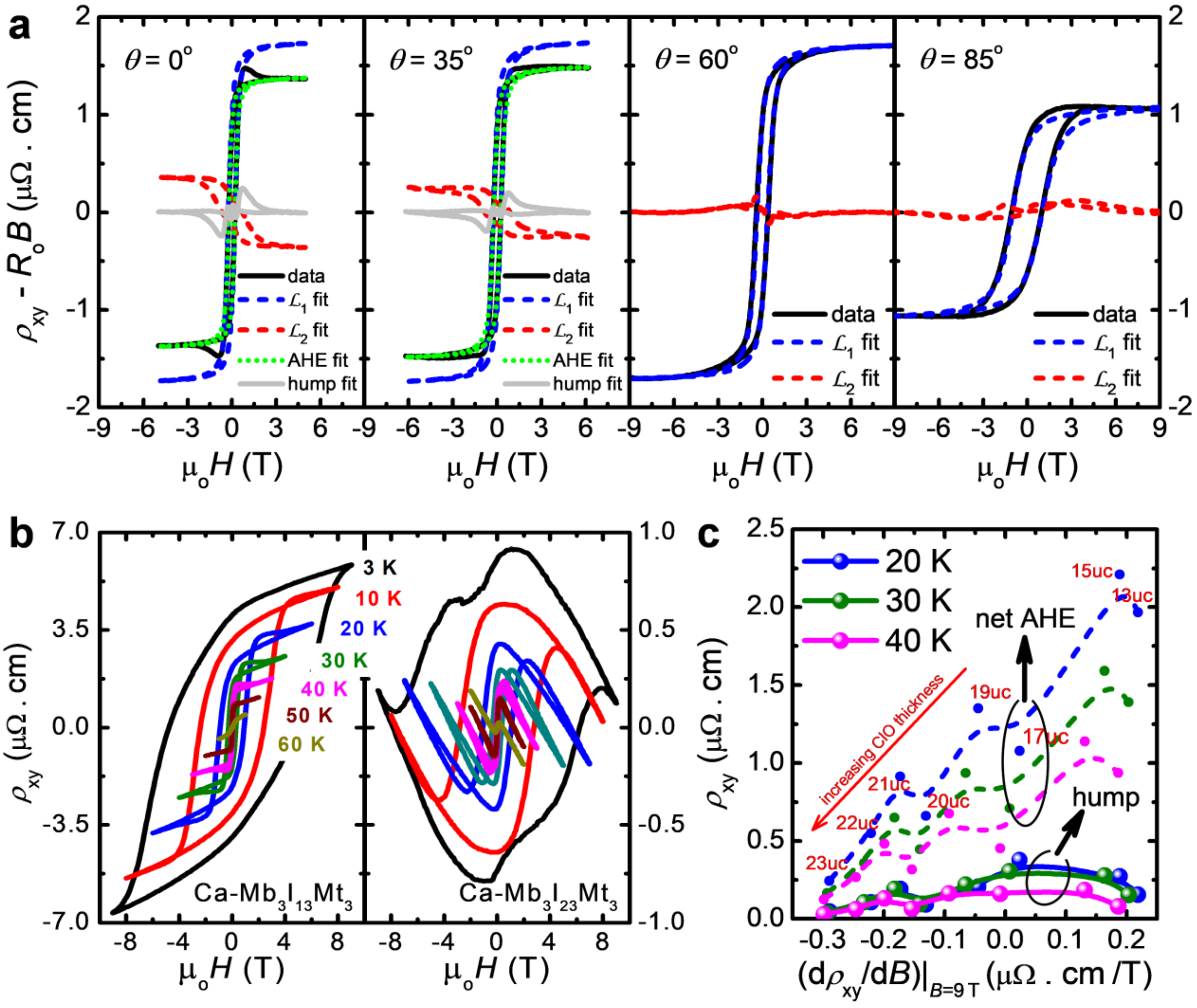

**Supporting Figure S1. (a)** Deconvolution of Extraordinary Hall Effect data into various components following the "AHE+hump" and "Bi-AHE" interpretations, measured at increasing θ-angle away from out-of-plane magnetic field. **(b)** Hall Effect (inclusive of the Ordinary Hall (OHE) component from Lorentz force) of the trilayer structure with different CIO spacer thicknesses. Note the OHE gradient sign-reversal accompanies the reduction of overall abscissa scale (right side), although the hump signal is present in both structures. **(c)** A summary of Extraordinary Hall Effect evolution presented in the "AHE+hump" format, with the tuning of CIO layer thickness and fixed CMO thickness at Mt=Mb=3uc.

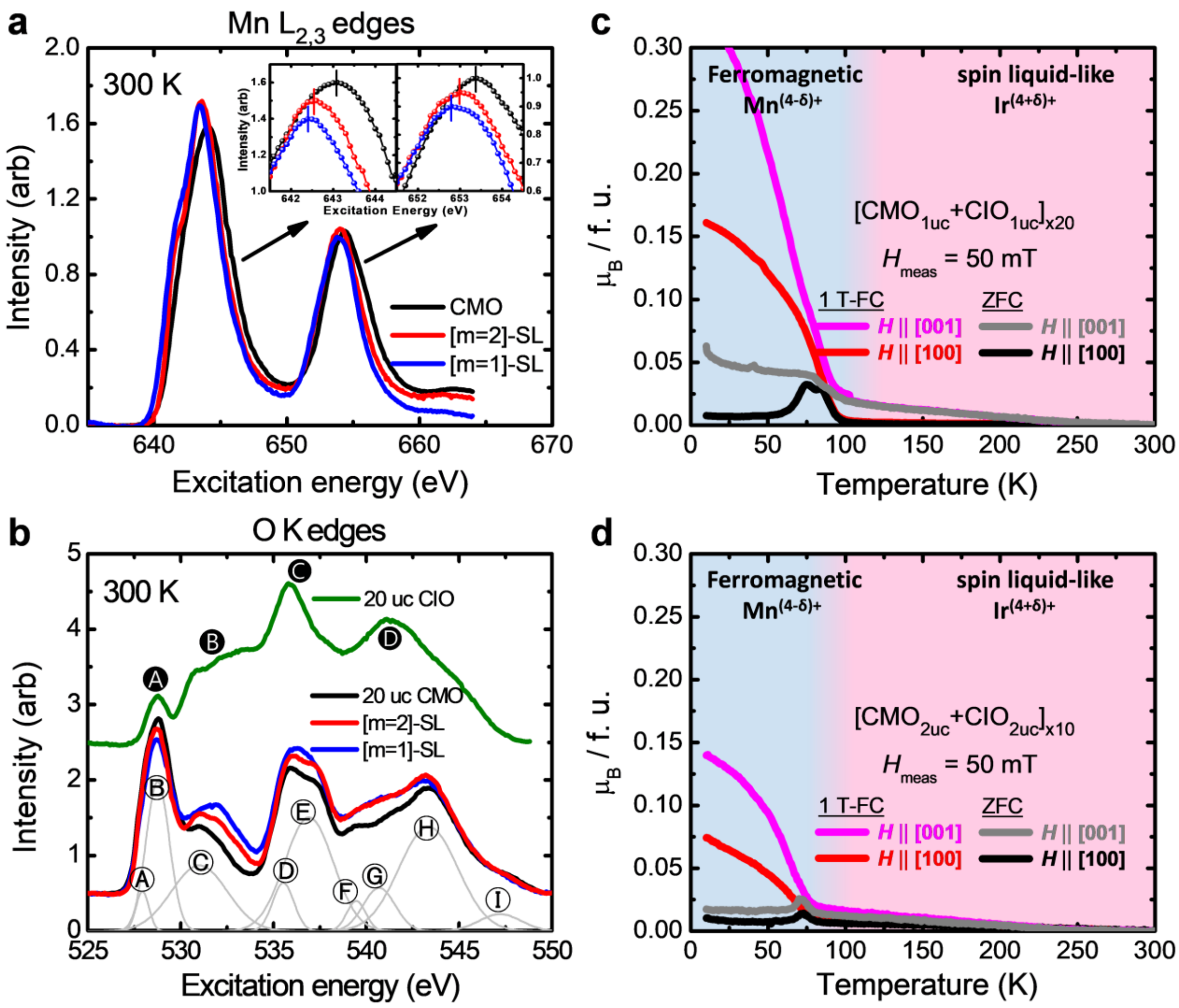

**Supporting Figure S2.** XAS spectra for **(a)** Mn $L_{2,3}$ edges and **(b)** O K edges of the thick $CaMnO_3$ film and [m=1,2]-SLs grown on LAO(001) with fitting peaks labelled Ⓐ to Ⓘ. In (b), the O K edges of thick CIO//LAO(001) is also shown for comparison. The baselines of the CIO, CMO and superlattices are shifted vertically away from the fitting peaks for clarity. M-T curves measured with in-plane and out-of-plane fields for the **(c)** [m=1]-SL and **(d)** [m=2]-SL.

a

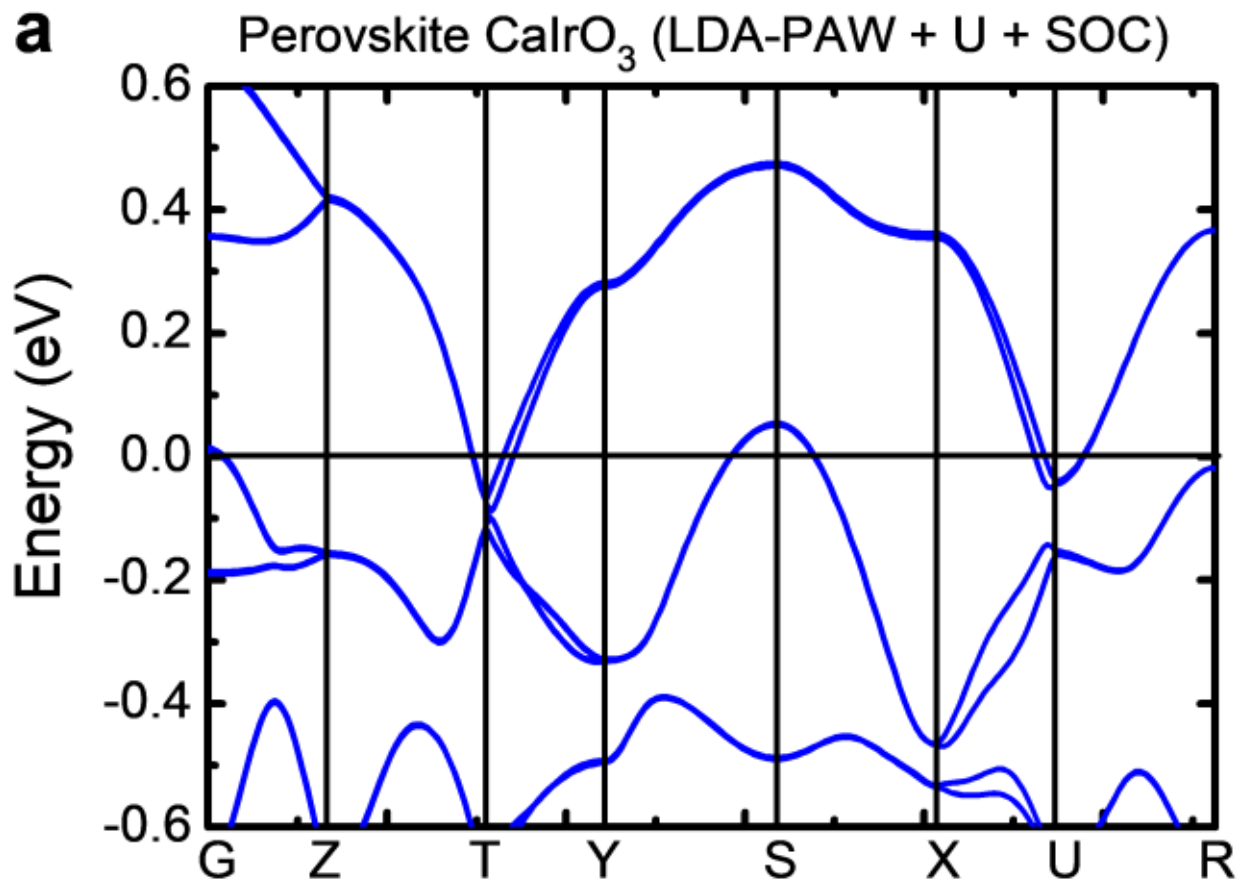

b

| Direction | Lattice parameter (Å) |
| --- | --- |
| a | 5.602 |
| b | 5.194 |
| c | 7.537 |

c

| Atom | $M_{tot}$ ($\mu_B$) | Bader charge |
| --- | --- | --- |
| Ir-1 | 0 | 7.40 |
| Ir-2 | 0 | 7.42 |
| Ir-3 | 0 | 7.40 |
| Ir-4 | 0 | 7.39 |

d

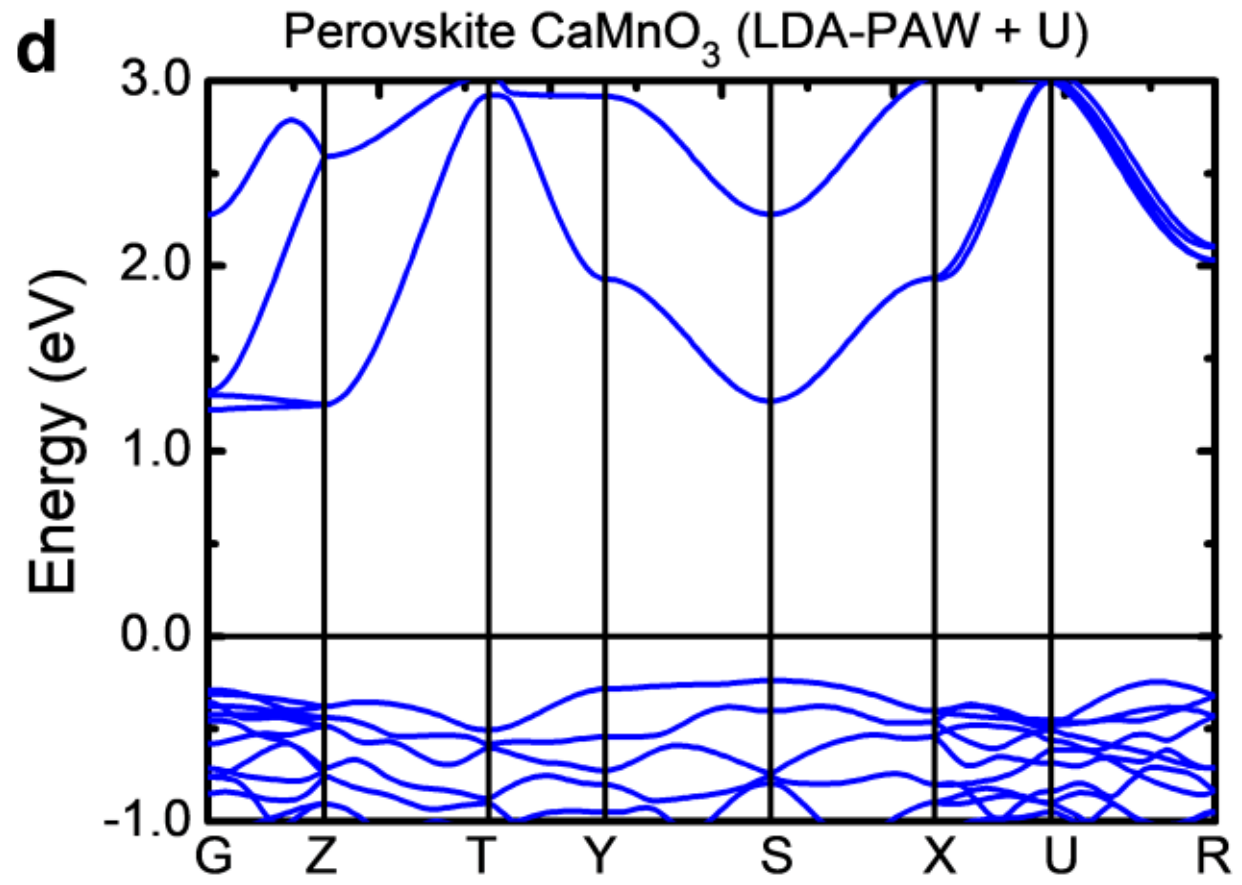

e

| Direction | Lattice parameter (Å) |
| --- | --- |
| a | 5.229 |
| b | 5.175 |
| c | 7.347 |

f

| Atom | $M_{tot}$ ($\mu_B$) | Bader charge |
| --- | --- | --- |
| Mn-1 | +2.712 | 11.11 |
| Mn-2 | -2.712 | 11.10 |
| Mn-3 | +2.712 | 11.11 |
| Mn-4 | -2.712 | 11.12 |

**Supporting Figure S3.** DFT calculations result for the bulk $CaIrO_3$ and $CaMnO_3$. **(a,d)** The E-k band diagrams of $CaIrO_3$ (a) and $CaMnO_3$ (d). **(b,e)** Orthorhombic lattice parameters for $CaIrO_3$ (b) and $CaMnO_3$ (e). **(c,f)** Moments and Bader charges for neighboring Ir (c) and Mn (f) species. The +/- signs for $CaMnO_3$ indicates G-type antiferromagnetic.

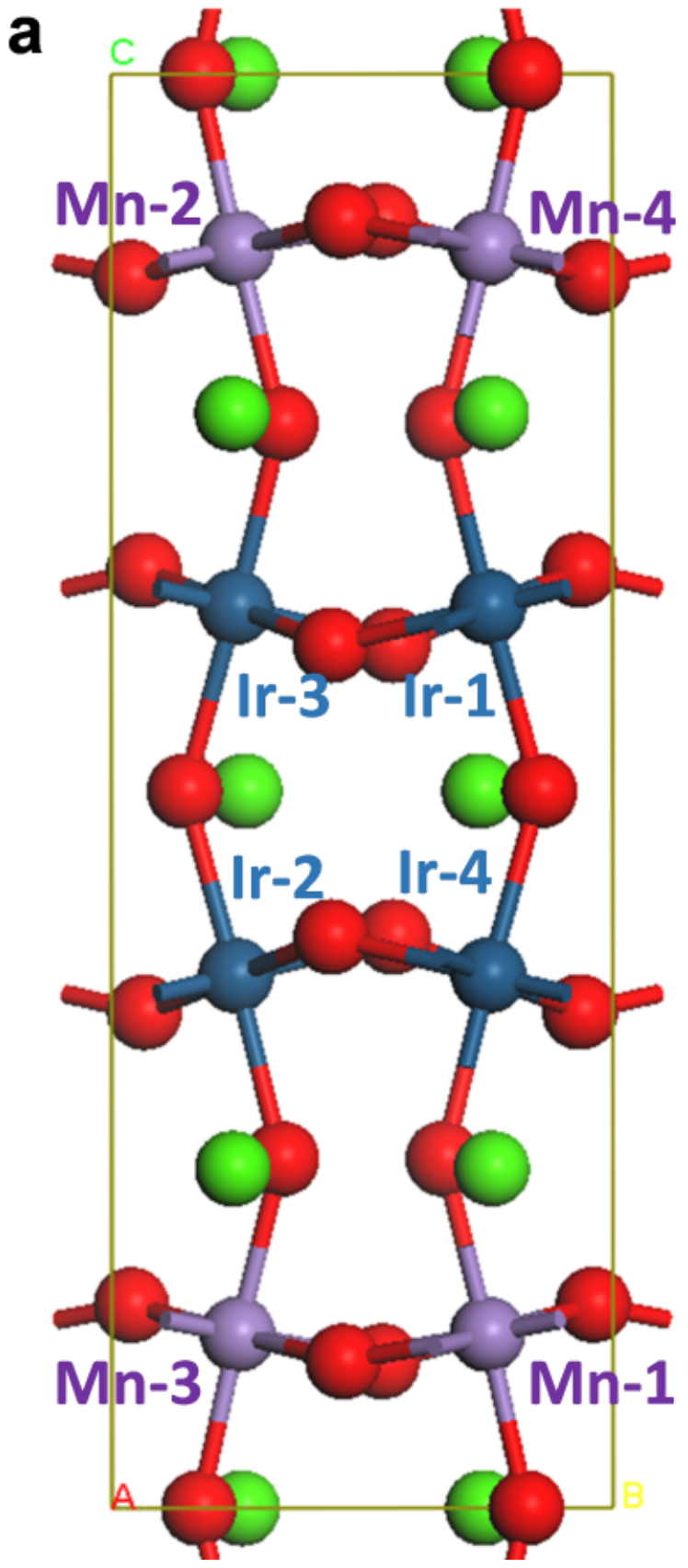

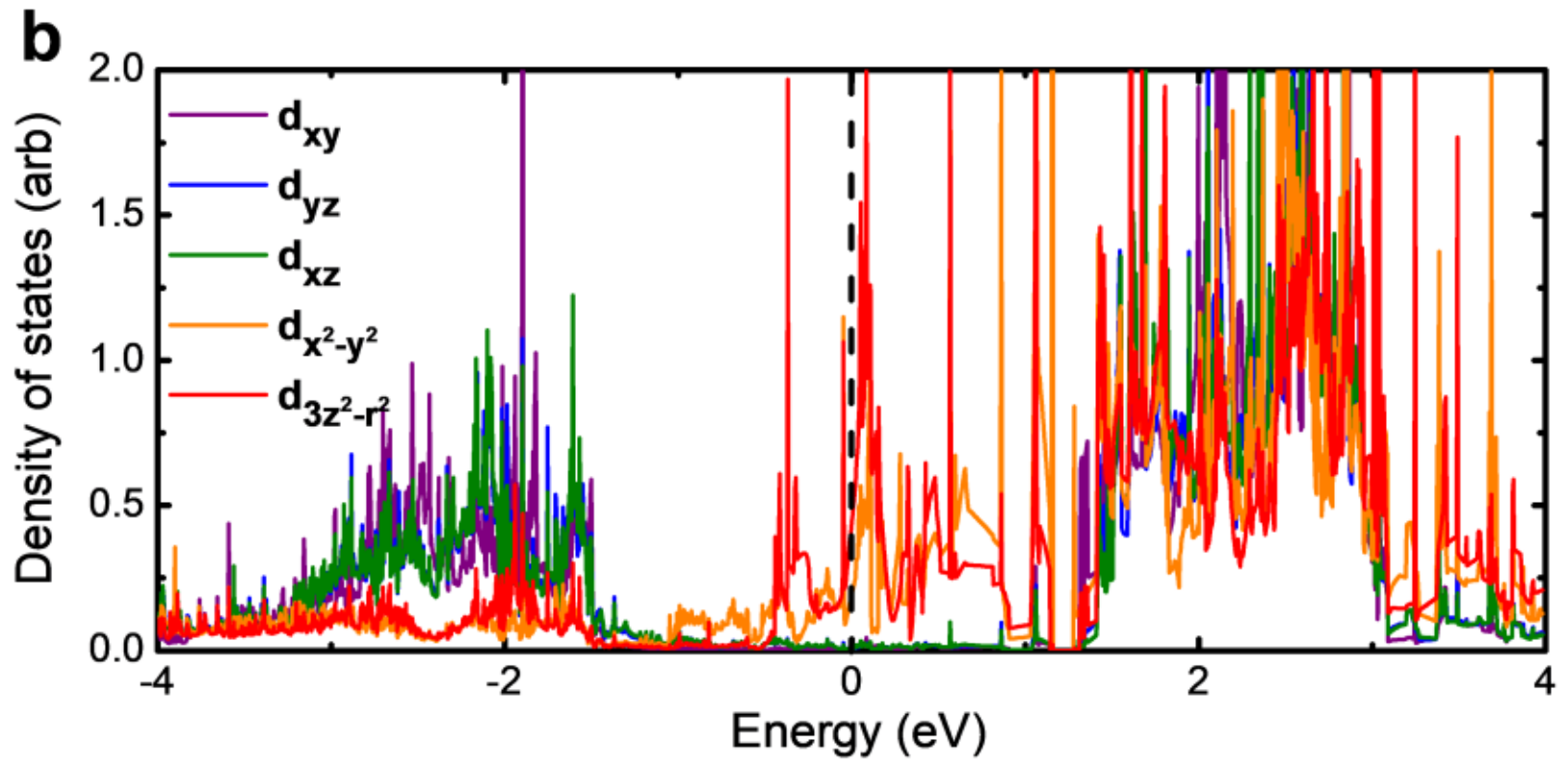

c

| Atom | Magnetic moment ($\mu_B$) | | | | Bader charge |
|---|---|---|---|---|---|
| | $M_x$ | $M_y$ | $M_z$ | $M_{tot}$ | |
| Mn-1 | 1.595 | 1.821 | 1.737 | 2.979 | 11.22 |
| Mn-2 | 1.642 | 1.819 | 1.695 | 2.980 | 11.24 |
| Mn-3 | 1.758 | 1.655 | 1.745 | 2.979 | 11.23 |
| Mn-4 | 1.806 | 1.676 | 1.674 | 2.979 | 11.22 |
| Ir-1 | -0.014 | -0.011 | -0.009 | 0.020 | 7.29 |
| Ir-2 | -0.005 | -0.006 | -0.030 | 0.031 | 7.28 |
| Ir-3 | -0.011 | -0.014 | -0.009 | 0.020 | 7.28 |
| Ir-4 | -0.006 | -0.005 | -0.030 | 0.031 | 7.29 |

**Supporting Figure S4.** DFT calculations for the [$CaIrO_3$ (2uc)/$CaMnO_3$ (2uc)]$_{x\infty}$ superlattice. **(a)** One unit supercell after relaxation. **(b)** Orbital-resolved DOS, with $E_F$ = 0 eV (dashed line). **(c)** Moments and Bader charges for the Mn and Ir species, with the atoms labeled referring to (a).